\documentclass[a4paper,10pt,twoside]{cpc-hepnp}

\usepackage{multicol}
\usepackage{siunitx}
\usepackage{graphicx}
\usepackage{booktabs}
\usepackage{amssymb,bm,mathrsfs,bbm,amscd}
\usepackage[tbtags]{amsmath}
\usepackage{lastpage}
\usepackage[utf8]{inputenc}
\usepackage{CJK}
\usepackage[colorlinks=true,linkcolor=blue,urlcolor=blue,citecolor=blue]{hyperref}

\def\tauqu{\tau_{\rm qu}}
\def\tautk{\tau_{\rm tk}}
\def\taucs{\tau_{\rm cs}}
\def\taubs{\tau_{\rm bs}}
\DeclareSIUnit\Torr{Torr}

\begin{document}
\begin{CJK*}{UTF8}{gbsn}

\title{Experimental Study using Touschek Lifetime as Machine Status Flag in SSRF\thanks{Work supported by National Natural Science Foundation of China (11075198)}}

\author{%
      CHEN Zhi-Chu(陈之初)$^{1,2;1)}$\email{chenzhichu@sinap.ac.cn}%
\quad LENG Yong-Bin(冷用斌)$^{1,2;2)}$\email{lengyongbin@sinap.ac.cn, corresponding author.}%
\\
\quad YUAN Ren-Xian(袁任贤)$^{1,2}$
\quad YAN Ying-Bing(阎映炳)$^{1,2}$
\quad YU Lu-Yang(俞路阳)$^{1,2}$
}
\maketitle

\end{CJK*}

\address{%
$^1$ Shanghai Institute of Applied Physics, Chinese Academy of Sciences, Shanghai 201800, China\\
$^2$ Shanghai Synchrotron Radiation Facility, Chinese Academy of Sciences, Shanghai 201203, China\\
}

\begin{abstract}
The stabilities of the beam and machine have almost the highest priority in a modern light source. Although a lot of machine parameters could be used to represent the beam quality, there lacks a single one that could indicate the global information for the machine operators and accelerator physicists, recently. A new parameter has been studied for the last few years as a beam quality flag in Shanghai Synchrotron Radiation Facility~(SSRF). Calculations, simulations and detailed analysis of the real-time data from the storage ring had been made and interesting results had confirmed its feasibility.
\end{abstract}



\begin{multicols}{2}

\section{Introduction}

Beam quality is of great importance, especially for a light source which aims at providing stable synchrotron radiation for scientific research. A couple of machine parameters have been maturely used in most third generation storage rings around the world to indicate the beam status, e.g., the transverse beam size/emittance from a pinhole camera, the variance of the close orbit from the BPM system, etc. Other parameters such as the beam length/energy spread from a streak camera are also monitored in some facilities. However, monitoring a single parameter seems not enough to reflect the beam status while monitoring all of them simultaneously would eventually confuse the operators.

During the selection of the necessary parameters to be monitored, an economic proposal was believed to be competitive: to use the beam current to get some factor of the beam. The beam lifetime could interpret the beam status in some way, but it is related to the beam charge so no convenient reference is available to say if the beam is in good status. Further processes are still needed to make this proposal a feasible solution.

\subsection{Beam Lifetime}

A bunch containing $N$ charged particles~(electrons in most third-generation synchrotron radiation sources) in a storage ring decays due to a variety of mechanisms. Some of the non-trivial causes are: quantum lifetime (emission of synchrotron radiation), Coulomb scattering (elastic scattering on residual gas atoms), Bremsstrahlung (photon emission induced by residual gas atoms) and Touschek effect (electron-electron scattering).

The relative loss rate at a given time of the quantity of the beam defines the lifetime $\tau$:
\begin{equation}\label{eq:lifetime-def}
\frac1\tau \equiv -\frac{\dot N}N = -\frac{\dot Q}Q ,
\end{equation}
where $Q = eN$ is the charge of the beam.

The beam lifetimes due to the quantum character of synchrotron radiation, the Touschek effect, the elastic Coulomb scattering and inelastic bremsstrahlung between the electron beam and the pure nitrogen gas are given by\cite{Wiedemann:PAP:2007}
\begin{align}
\tauqu &= \frac12 \tau_w \frac{e^\xi}\xi , \label{eq:quantum-lifetime}\\
\frac1\tautk &= \frac{r_{\mathrm c}^2 c Q}{8 \pi e \sigma_x \sigma_y \sigma_\ell} \frac{\lambda^3}{\gamma^2} D(\epsilon) , \label{eq:Touschek-lifetime}\\
\taucs(\si{hours}) &= 10.25 \frac{(cp)^2(\si{\giga\electronvolt\squared})\epsilon_{\mathrm A}(\si{\milli\metre\milli\radian})}{\langle \beta (\si{\metre}) \rangle P(\si{\nano\Torr})} , \label{eq:Coulomb-scattering-lifetime}\\
\taubs^{-1}(\si{hours^{-1}}) &= 0.00653 P (\si{\nano\Torr}) \ln \frac1{\delta_{\mathrm{acc}}} , \label{eq:bremsstrahlung-lifetime}
\end{align}
where $\tau_w$ is the damping time, $\xi$ a function of the acceptance of the beam and the size of the beam bunch $\xi = \frac{A^2}{2\sigma^2}$, $r_{\mathrm c}$ the classical electron radius, $(\sigma_x,\sigma_y,\sigma_\ell)$ the three dimensions of the bunch, $\lambda^{-1} = \Delta p/p_0|_{\mathrm{rf}}$ the RF momentum acceptance of the ring, $P$ the pressure and $D(\epsilon)$ the Touschek lifetime function
\begin{align}
D(\epsilon) &= \sqrt\epsilon \Biggl[ -\frac32 e^{-\epsilon} + \frac\epsilon2 \int_\epsilon^\infty \frac{\ln u}u e^{-u} \,\mathrm du \nonumber\\
&+ \frac12(3\epsilon-\epsilon\ln\epsilon+2)\int_\epsilon^\infty \frac{e^{-u}}u \,\mathrm du \Biggr] \label{eq:Touschek-function}\\
\epsilon &= \left( \frac{\beta_x\Delta p_{\mathrm{rf}}}{mc\gamma^2\sigma_x} \right)^2 ,
\end{align}

The total beam loss rate is the sum of all kinds of beam loss rates contributed by individual beam loss mechanisms:
\begin{equation}\label{eq:lifetime-decomposition}
\frac1\tau = \frac1\tauqu + \frac1\tautk + \frac1\taucs + \frac1\taubs .
\end{equation}

\subsection{The Touschek Lifetime}

None of the components in the r.h.s.\ of \eqref{eq:lifetime-decomposition} is beam charge related except for the Touschek effect based on equations~\eqref{eq:quantum-lifetime}, \eqref{eq:Touschek-lifetime}, \eqref{eq:Coulomb-scattering-lifetime} and~\eqref{eq:bremsstrahlung-lifetime}. The Touschek lifetime is of great importance and has already been simulated\cite{Khan:EPAC:1994,Boscolo:PAC:2009:TH6PFP060} and measured\cite{Kang:APAC:2001,Leonov:RuPAC:2004,Steier:PAC:2009:TH5PFP033,Blinov:PAC:2011:MOP182,Nash:IPAC:2011:THPC008} in many light sources. It is proportional to the beam charge as shown in equation~\eqref{eq:Touschek-lifetime} so that the total beam loss rate can be simplified as the following equation if we use a ``Touschek lifetime factor'' $k$ to represent the Touschek lifetime:
\begin{equation}\label{eq:tau-vs-q}
\frac1\tau = \frac1{\tau_0} + kQ ,
\end{equation}
where $\tau_0$ is the combined quantum and vacuum lifetime
\begin{equation}
\frac1{\tau_0} = \frac1\tauqu + \frac1\taucs + \frac1\taubs ,
\end{equation}
and
\begin{equation}
k = \frac{r_{\mathrm c}^2 c}{8 \pi e \sigma_x \sigma_y \sigma_\ell} \frac{\lambda^3}{\gamma^2} D(\epsilon) ,
\end{equation}
so that
\begin{equation}
\tautk = \frac1{kQ} .
\end{equation}

For a quasi-steady state---e.g., the magnets, vacuum level, RF voltage, tunes and other machine parameters remain unchanged within a period of time, which is almost exclusive in many storage rings---the quantum lifetime, vacuum lifetime and the Touschek lifetime factor can all be considered as constants. A differential function about $Q$ and $t$ can be derived from equations~\eqref{eq:lifetime-def} and~\eqref{eq:tau-vs-q}:
\begin{equation}\label{eq:q-diff-formula}
-\frac{\dot Q}Q = \frac1{\tau_0} + kQ .
\end{equation}
Hence
\begin{equation}\label{eq:current-formula}
\frac1Q = k\tau_0 \left[ \exp\left(\frac{t-t_0}{\tau_0}\right) - 1 \right] ,
\end{equation}
where
\begin{equation}
t_0 = \tau_0 \ln \frac{k\tau_0 Q(0)}{k\tau_0 Q(0) + 1}
\end{equation}
is a constant of integration. Thus, a perfect machine would have an almost exponentially decreasing current curve.

As can be easily observed, the Touschek factor $k$ is inversely proportional to the beam volume $\sigma_x \sigma_y \sigma_\ell$, the square of the beam energy and the cube of the momentum acceptance. The slope of the function $D(\epsilon)$ is negligible in the field of small $\epsilon$, i.e., high energy when observing the deviate of the r.h.s.\ of equation~\eqref{eq:Touschek-function}. If $D(\epsilon)$ is regarded as a constant, the relative Touschek factor can be easily determined by a simple algebraic form of the beam volume, beam energy and the RF acceptance.

\section{Critical Factors in the Measurements}

As a practical system, the beam diagnostics in SSRF could not remove the measurement error totally. Besides, the physical variables mentioned above have hardly no disturbances. Although $\tau_0$ and $k$ are treated as invariants in~\eqref{eq:q-diff-formula}, it may still be necessary to estimate the model performance.

\subsection{Impact of the Quantum Lifetime}

The transverse acceptances are no less than \SI{3}{\milli\metre} in SSRF, and the transverse beam sizes are about \SI{80}{\micro\metre}/\SI{20}{\micro\metre}. The transverse damping time is \SI{1.3}{\milli\second}. The quantum lifetime in \eqref{eq:quantum-lifetime} is regarded long enough in SSRF, as well as in other third generation light sources. The fluctuation of the quantum lifetime could barely touch the total lifetime.

\subsection{Influence of the Pressure}

The vacuum pressure is detected not constant during the operation and should not be ignored in a precision system. Equations~\ref{eq:Coulomb-scattering-lifetime} and~\eqref{eq:bremsstrahlung-lifetime} show that the beam loss rate of either Coulomb scattering or bremsstrahlung effect is directly proportional to the pressure. Thus, equation~\eqref{eq:tau-vs-q} should include the pressure related part to extract the Touschek lifetime:
\begin{equation}\label{eq:tau-vs-q-and-p}
\frac1\tau = \frac1\tauqu + mP + kQ .
\end{equation}

The resolution and accuracy of the vacuum gauge might not be satisfying in this situation, but the reading of the gauge $P_1=P+P_0+n(P)$ would not be much trouble. The constant offset $P_0$ could be included in the quantum lifetime part which is never paid attention to. The noise $n(P)$ can be decreased by curve fitting.

\subsection{Contribution of the Beam Size Shift}

The parameters which can be easily measured in a storage ring are the sizes of the beam. The beam length $\sigma_\ell$ measurement would involve a streak camera, and both the precision and the update speed were not satisfying right now. The non-linearities of the screen or the camera had already been calibrated carefully so that we would use the transverse beam sizes as a comparison and an aid in our data analysis. The sizes which were calculated by using the original X-ray image of the radiation\cite{Huang:NT:2010} and the point spread functions~(PSF)\cite{Chen:IPAC:2013:MOPME053} may have baseline offsets due to the measurement errors of the PSFs. This could have some effects to the Touschek lifetime fitting with the transverse sizes changing. We can expand the $\sigma_i$~($i$ is $x$ or $y$) terms in Taylor's series based on equation~\eqref{eq:Touschek-lifetime}:
\begin{equation}
\frac1{\sigma_{0,i}} = \frac1{\sigma_i} \left( 1-\frac{\Delta_{{\mathrm{PSF}},i}}{\sigma_i^2} \right)^{-\frac12}
= \sum_{n=0}^\infty \frac{(2n)!}{(2^nn!)^2} \frac{\Delta_{{\mathrm{PSF}},i}^n}{\sigma_i^{2n+1}},
\end{equation}
where $\sigma_{0,i}$ is the actual beam size, $\sigma_i$ is the calculated beam size by using the measured profile size $\sigma_{\gamma,i}$, the calibrated PSF $\sigma_{{\mathrm{PSF}},i}$ and the relation $\sigma_i^2 = \sigma_{\gamma,i}^2 - \sigma_{{\mathrm{PSF}},i}^2$, $\Delta_{{\mathrm{PSF}},i}$ is the difference between the square of the real PSF and the square of the measured one. The Touschek lifetime then can be expressed as the following form:
\begin{equation}
\frac1\tautk = AQ \frac1{\sigma_x}\left(1+\sum_k \frac{a_k}{\sigma_x^{2k}}\right) \frac1{\sigma_y}\left(1+\sum_k \frac{b_k}{\sigma_y^{2k}}\right) .
\end{equation}
If $\Delta_{{\mathrm{PSF}},i}$ is relatively small by comparing to $\sigma_i^2$, which is hopefully the truth, the higher order terms of the r.h.s.\ could be omitted.

\section{Data Analysis}

The data are being recorded recursively without interfering the operations of the machine as part of the global data warehouse system.\cite{Leng:IPAC:2011:TUPC116} An analysis has been made before anything should go on-line.

\subsection{Lifetime Calculation}

As much as we would like to use equation~\eqref{eq:current-formula} to get the Touschek factor, there were still ? problems. First of all, equation~\eqref{eq:current-formula} cannot be linearized which will make the fitting a little complex. Nevertheless, the propagation of the fitting errors of other parameters would certainly affect the accuracy and resolution of the interested factor.

A polynomial regression based algorithm has been used to calculate the beam lifetime. A reasonable period has been chosen to be polynomially fitted:
\begin{equation}
I_{n\times1} \simeq X_{n\times(k+1)} A_{(k+1)\times 1} ,
\end{equation}
where $I$ is the beam current vector, $X$ the time matrix and $A$ the coefficient vector: $I=(I_1, I_2, \dots, I_n )^T$, $A=(A_0, A_1, \dots, A_k)^T$ and
\begin{equation*}
X =
\begin{pmatrix}
1 & t_1 & \cdots & t_1^k \\
1 & t_2 & \cdots & t_2^k \\
\vdots & \vdots & \ddots & \vdots \\
1 & t_n & \cdots & t_n^k \\
\end{pmatrix} .
\end{equation*}
Thus the least-mean-square solution of the coefficient matrix is $A=(X^TX)^{-1}X^TI$, and the derivate would be
\begin{equation}
\dot I_{n\times1} \simeq X_{n\times k}^{(1)}A_{k\times 1}^{(1)} ,
\end{equation}
where $\dot I=(\dot I_1, \dot I_2, \dots, \dot I_n)^T$, $A^{(1)}=(A_1, A_2, \dots, A_k)^T$ and
\begin{equation*}
X^{(1)} =
\begin{pmatrix}
1 & t_1 & \cdots & t_1^{k-1} \\
1 & t_2 & \cdots & t_2^{k-1} \\
\vdots & \vdots & \ddots & \vdots \\
1 & t_n & \cdots & t_n^{k-1} \\
\end{pmatrix} \cdot \mathop{\mathrm{diag}}(1,2,\dots,k) .
\end{equation*}
Therefore, the lifetime
\begin{equation}
\frac1\tau = -\frac{\dot Q}{Q} = -\frac{\dot I}{I}
\end{equation}
can be calculated by using the beam current data. A further weighted average process is needed to decrease the current noise and fitting errors by using overlapped intervals to estimate the lifetimes.

The calculated lifetime is actually the weighted averaged lifetime of the filled bunches. Since the bunches were evenly filled, this averaged lifetime could be regarded as the lifetime of each bunch.

\subsection{Vacuum Lifetime Estimate}

The ingredient of the gas in the vacuum chamber is assumed to be invariant and the vacuum lifetime is inversely proportional to the pressure $P$. If the pressure varies while every other parameter related to the lifetime holds its own value, the relation between the pressure and the corresponding vacuum lifetime with respect to the specific gas ingredient in SSRF can be easily calibrated. Fortunately, the pressure experienced a long significant change due to the destruction of the vacuum after an upgrade of the storage ring.

A series of data was carefully chosen to ensure they share the same current and the same transverse size, which presumes that the Touschek lifetime is fixed and can be regarded as a constant. So the total lifetime is linearly related to the pressure: $\tau^{-1} = AP+b$ where $A$ and $B$ are the coefficients to be determined (as shown in figure~\ref{fig:vacuum-lifetime-formula}).

\begin{center}\begin{minipage}{\hsize}
\includegraphics[width=.8\hsize]{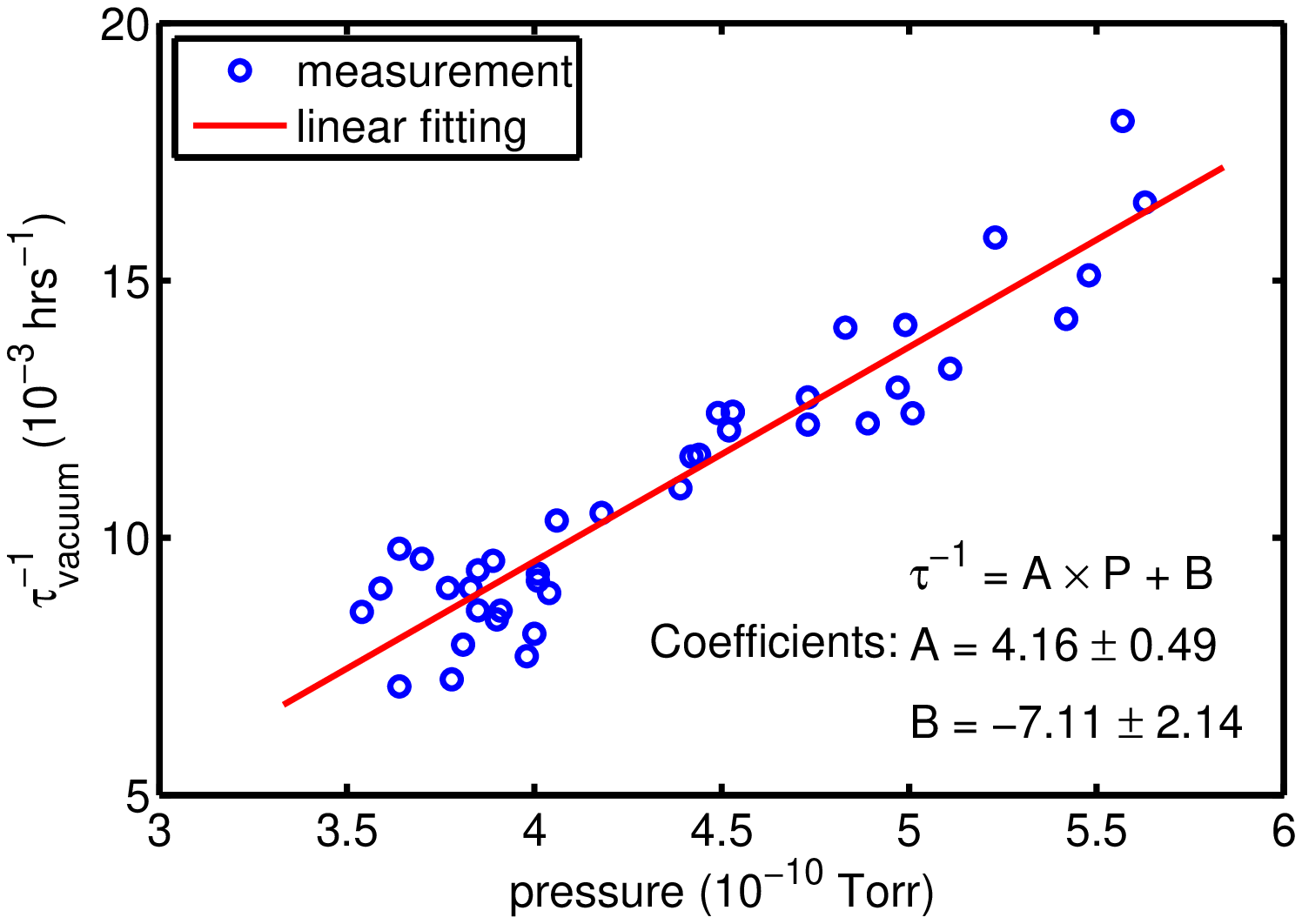}
\figcaption{\label{fig:vacuum-lifetime-formula}Estimate the practical form of the vacuum lifetime and the pressure. The data were tracked after a leakage of the storage ring during a hardware upgrade in 2012.}
\end{minipage}\end{center}

\subsection{Touschek Lifetime as a Beam Quality Factor}

Ignoring the quantum lifetime, the Touschek lifetime $\tautk$ would then be separated from the total lifetime by using the real time average pressure data $P$ provide by the vacuum gauges distributed around the storage ring and the coefficient $A$ in figure~\ref{fig:vacuum-lifetime-formula} to eliminate the vacuum lifetime part. The Touschek factor $k=1/Q\tautk$ would be calculated afterwards.

The transverse beam sizes, where were calculated by using an X-ray pinhole image system, had been used as an aid to diagnose the ability of the Touschek factor. An illustration of the relation between the beam size and the Touschek factor is shown in figures~\ref{fig:TouschekFactor-TransverseArea-NormalDecay-Oct15} and~\ref{fig:TouschekFactor-TransverseArea-NormalDecay-Oct15-fitting}. The results had shown a strong linear correlation between the factor and the reciprocal of the beam size in a normal smooth operation period, as expected.

\begin{center}\begin{minipage}{\hsize}
\includegraphics[width=.8\hsize]{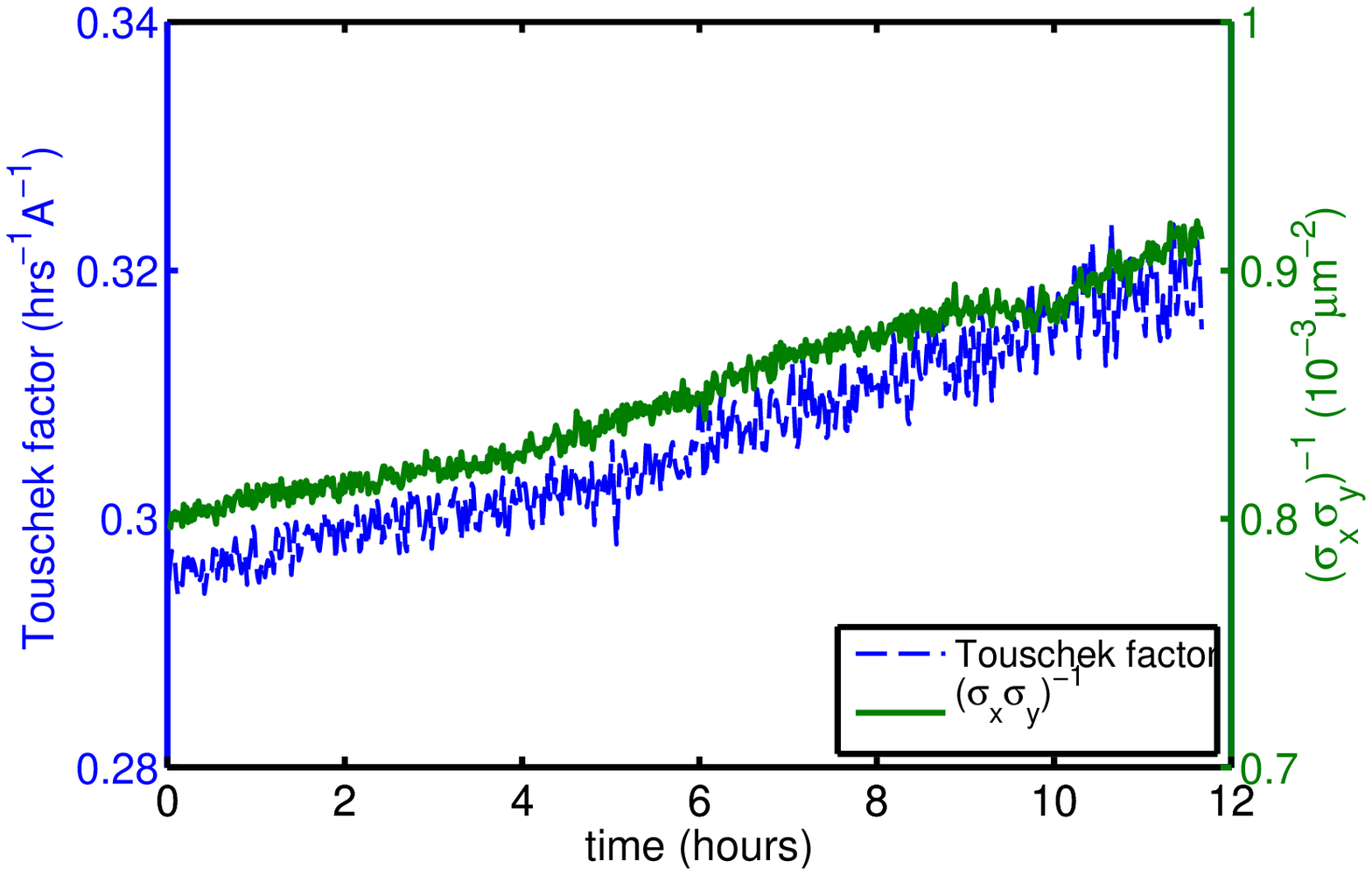}
\figcaption{\label{fig:TouschekFactor-TransverseArea-NormalDecay-Oct15}A typical Touschek factor and beam size trend during a successful decay period. Data were acquired at Oct.~15.}
\end{minipage}\end{center}

\begin{center}\begin{minipage}{\hsize}
\includegraphics[width=.8\hsize]{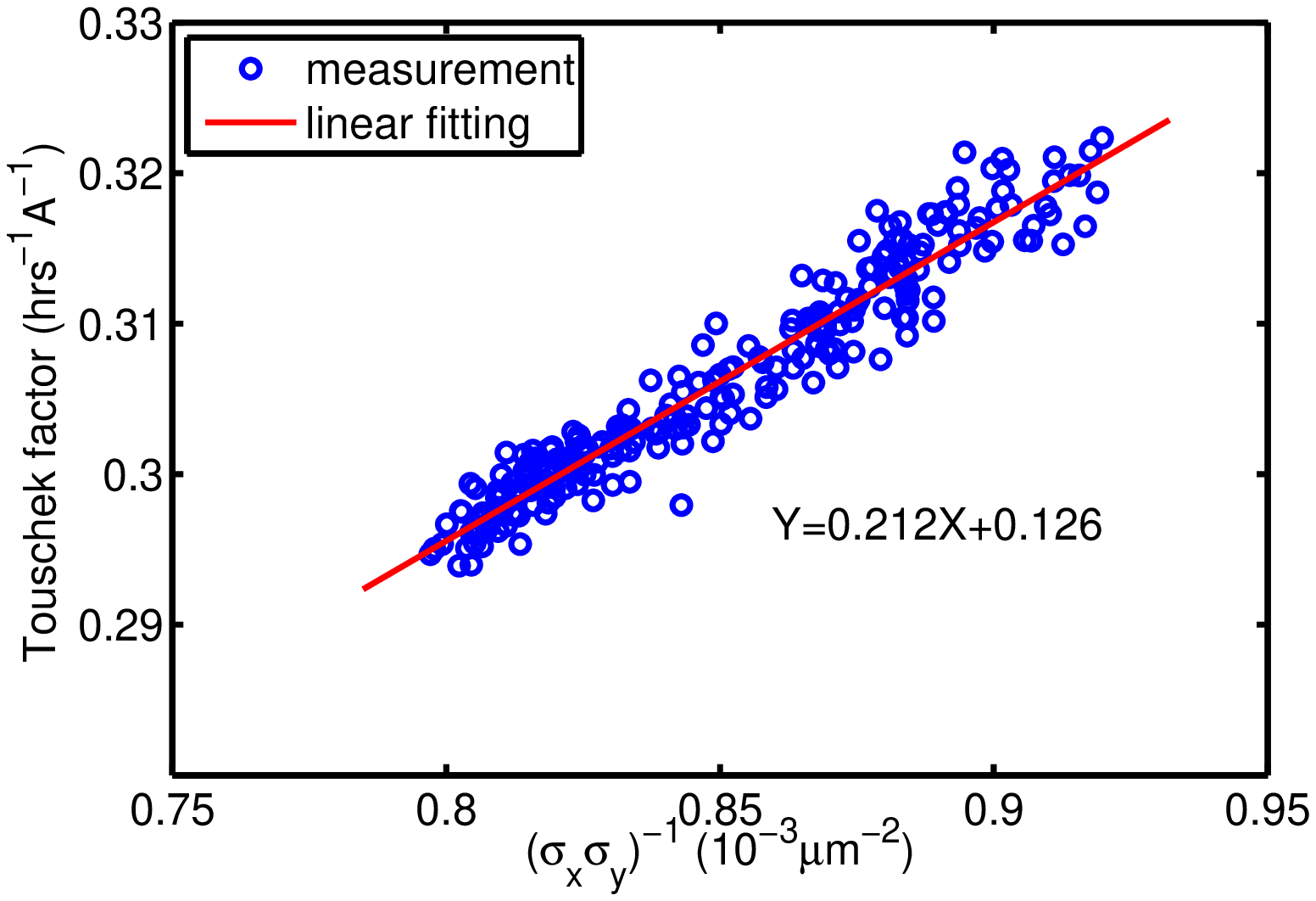}
\figcaption{\label{fig:TouschekFactor-TransverseArea-NormalDecay-Oct15-fitting}The strong linear relation between the Touschek factor and the beam size.}
\end{minipage}\end{center}

A sudden change of the beam size, which always indicates a change of the lattice or other configuration of the machine, is not desirable during the operations and need extra attentions. Figure~\ref{fig:TouschekFactor-TransverseArea-SingleDecay} demonstrates that the Touschek factor responded rapidly to the sudden change of the beam size.

\begin{center}\begin{minipage}{\hsize}
\includegraphics[width=.8\hsize]{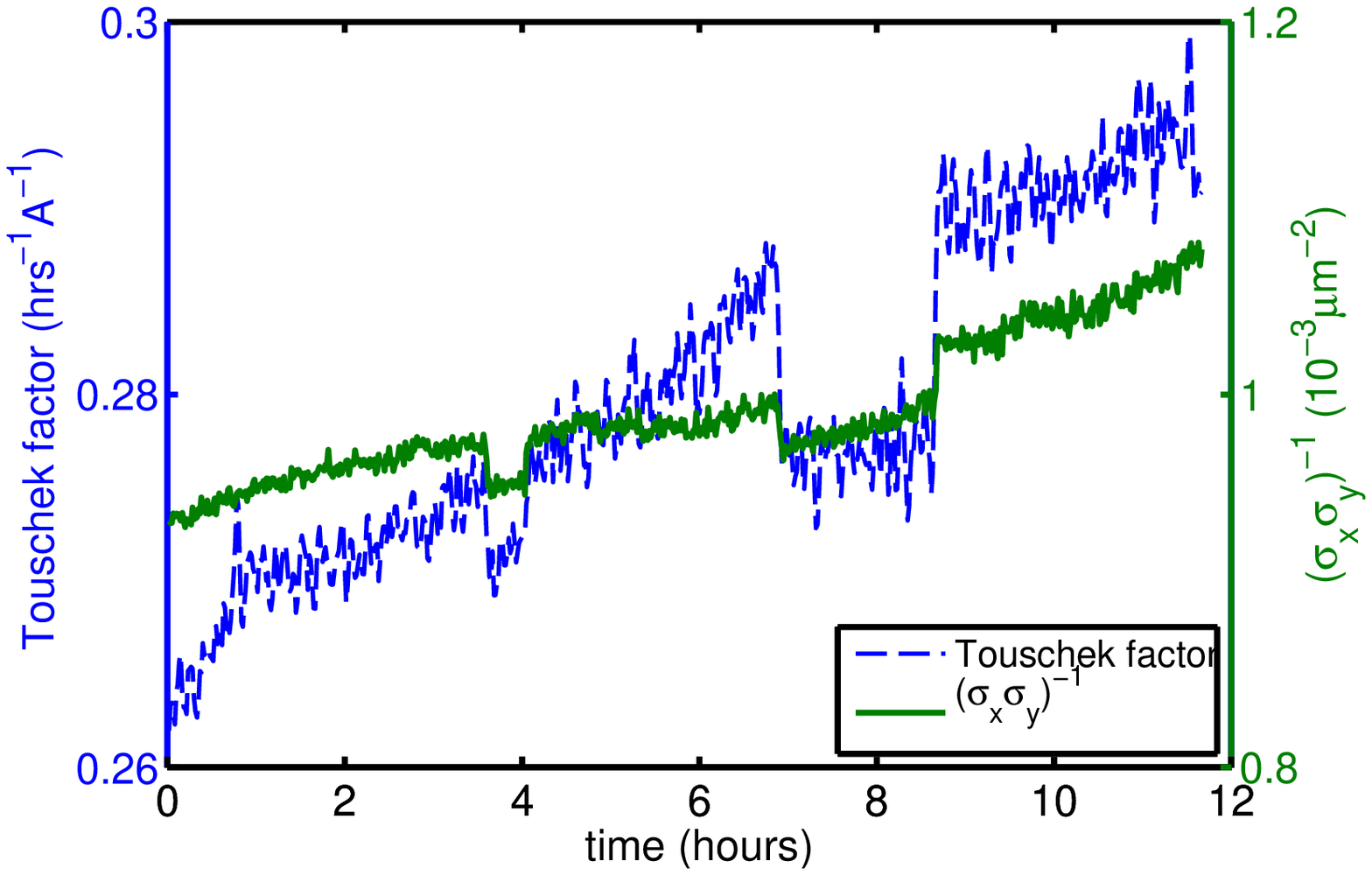}
\figcaption{\label{fig:TouschekFactor-TransverseArea-SingleDecay}Fast response to the sudden change of the beam size of the Touschek factor.}
\end{minipage}\end{center}

Someone might have noticed that, in figure~\ref{fig:TouschekFactor-TransverseArea-SingleDecay}, although the trends of the beam size and the Touschek factor are similar, the relative amplitude and the slope of the Touschek factor have some different information involved. This is because the Touschek factor is a global parameter of the machine and it is not just affected by the beam size.

Figures~\ref{fig:TouschekFactor-TransverseArea-Abnormal-Oct24} and~\ref{fig:TouschekFactor-TransverseArea-Abnormal-Oct24-analysis} gave a more detailed illustration. Since it was in the decay mode, figure~\ref{fig:TouschekFactor-TransverseArea-Abnormal-Oct24} should be viewed from right to left. There was a threshold at the beam current of \SI{152}{\milli\ampere} in the Touschek factor curve but the abnormality of the beam size was not quite visible without further analysis. The machine status was clearly divided into two different groups, as shown in figure~\ref{fig:TouschekFactor-TransverseArea-Abnormal-Oct24-analysis}.

\begin{center}\begin{minipage}{\hsize}
\includegraphics[width=.8\hsize]{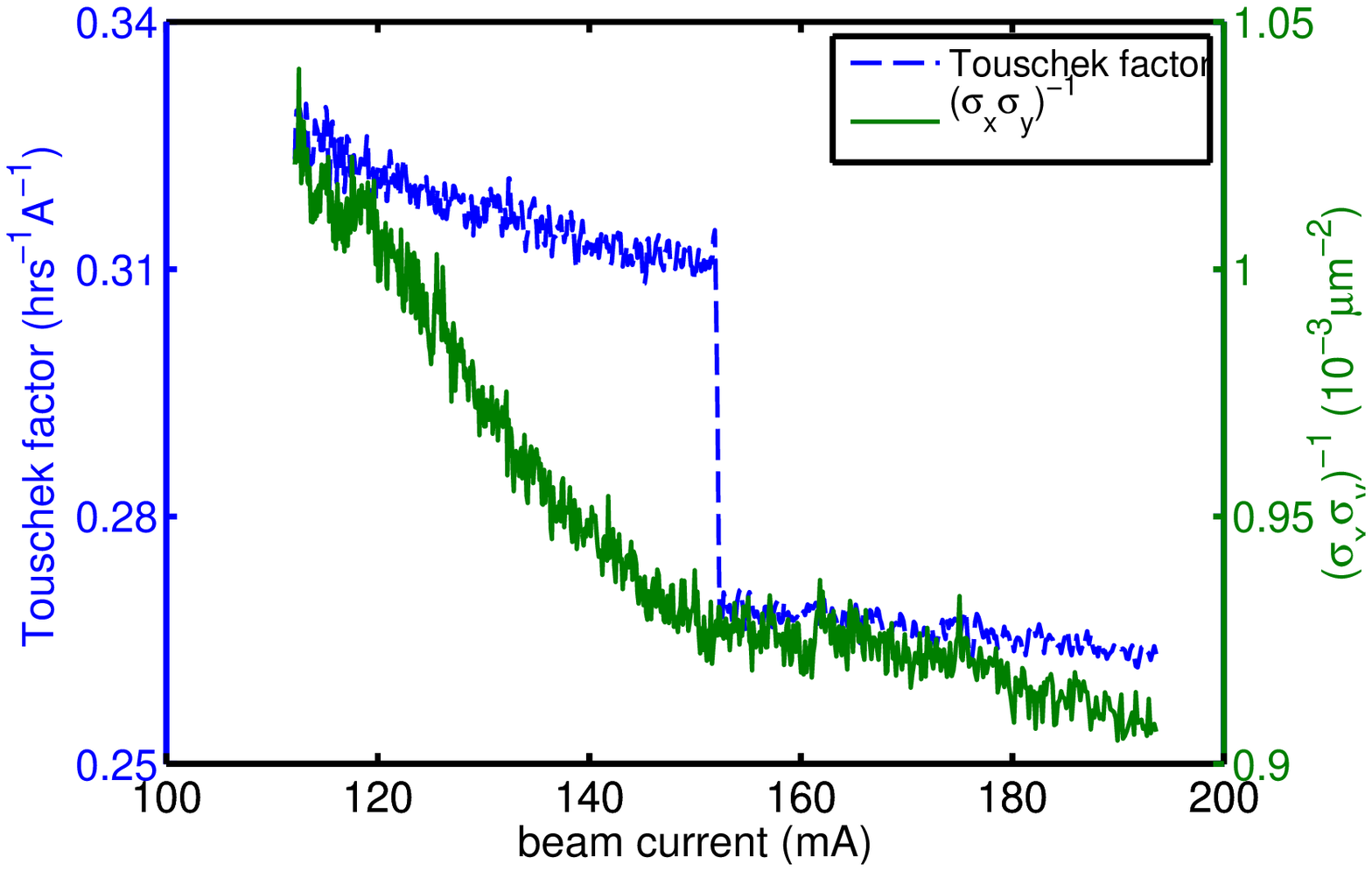}
\figcaption{\label{fig:TouschekFactor-TransverseArea-Abnormal-Oct24}Fast response of the Touschek factor to the sudden change of the machine status, while the beam size didn't notice the difference. Data were acquired at Oct.~24.}
\end{minipage}\end{center}

\begin{center}\begin{minipage}{\hsize}
\includegraphics[width=.8\hsize]{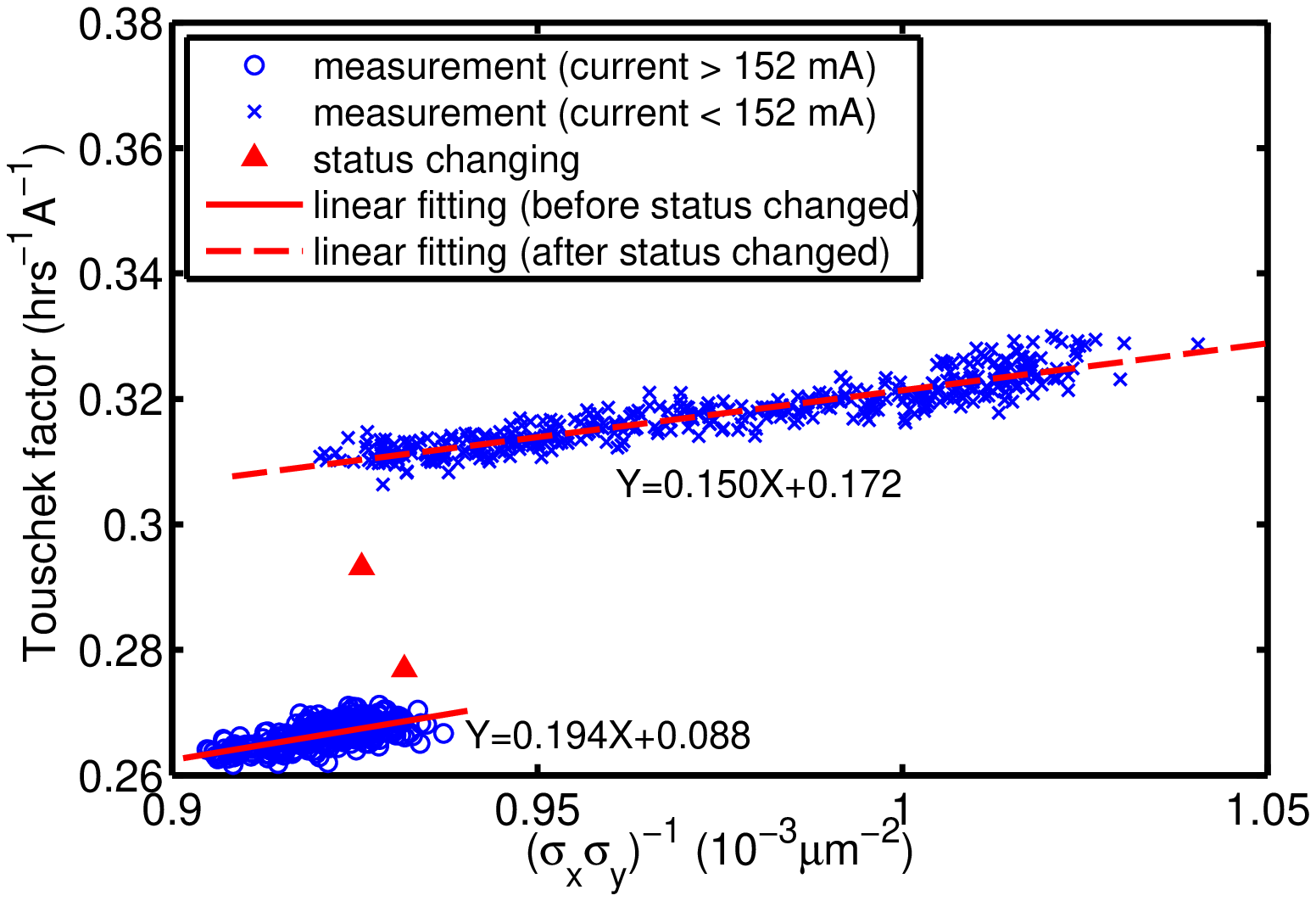}
\figcaption{\label{fig:TouschekFactor-TransverseArea-Abnormal-Oct24-analysis}The data are categorized into two group with the visible intermediate state.}
\end{minipage}\end{center}

This separation implies that the configuration could have been changed at the specific time or there was a specific mechanism that would introduce a new physical mode and the beam quality could deteriorate whenever the beam current is less than \SI{152}{\milli\ampere}. Two reasonable explanations had been made before further investigations: the change of the gap of an undulator, or the nonlinear effect of the machine.

If the Touschek factor has the ability to indicate the beam quality all by itself, different beam statuses could be able to be separated and categorized. Figure~\ref{fig:TouschekFactor-TransverseArea} shows another period of operations which had been interrupted three times for various reasons. The beam status was believed to be stable during each successful piece. The beam size might be continuous and inseparable if we ignore the second piece. The Touschek factor, on the other hand, gave significant jumps between each piece.

\begin{center}\begin{minipage}{\hsize}
\includegraphics[width=.8\hsize]{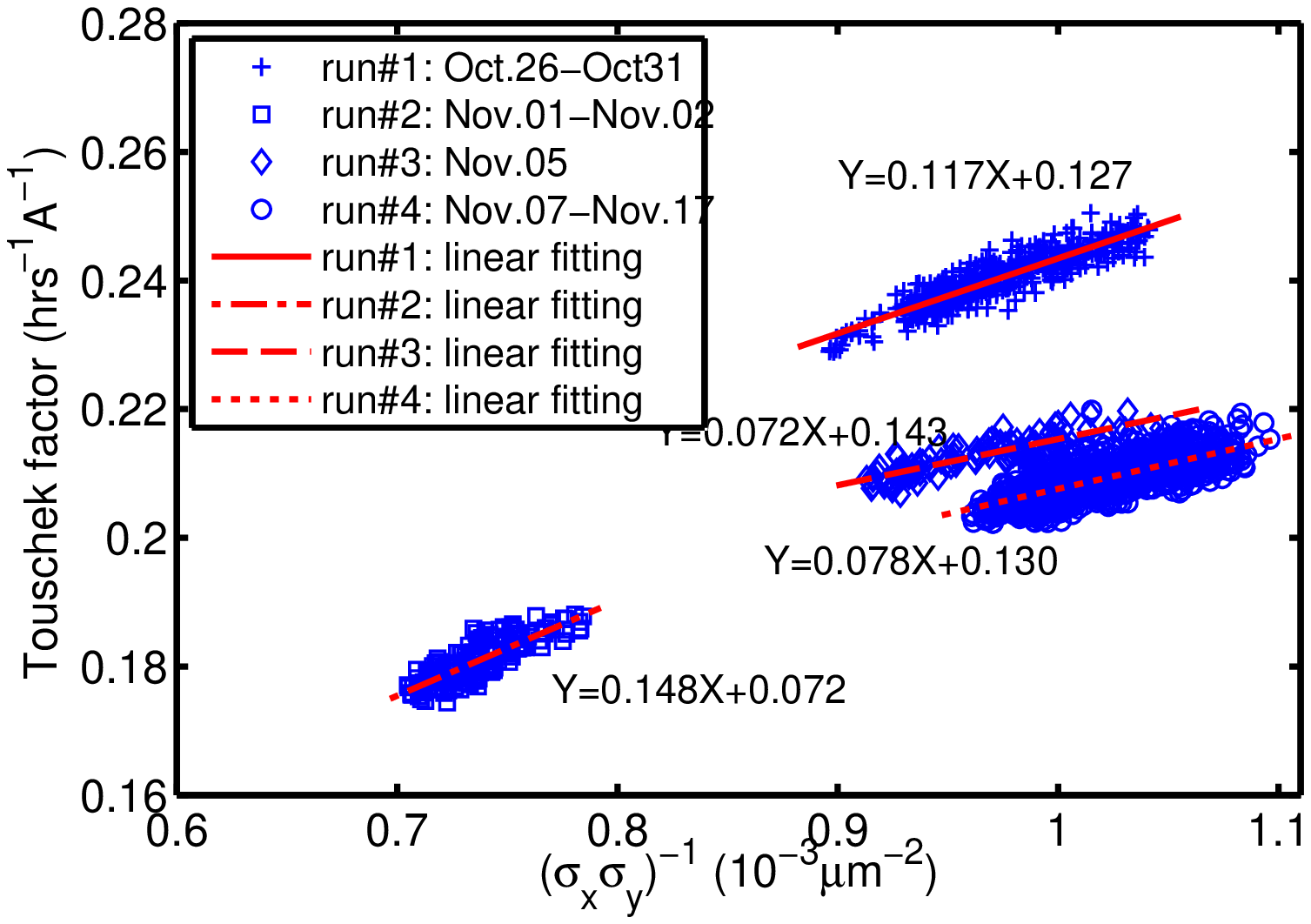}
\figcaption{\label{fig:TouschekFactor-TransverseArea}Different operation period belongs to different groups with different beam/machine status.}
\end{minipage}\end{center}

The beam size change seen at the pinhole camera is local and should be cross checked with the $\beta$-function and the bunch length, which was quite stable during the operations, to confirm a change of emittance. But the Touschek factor alone is able to diagnose the machine status. Since some of the machine parameters, such as the bunch length, dispersion functions and $\beta$ functions, are difficult to be monitored on-line, this Touschek factor therefore

\section{Conclusions}

In order to find a global flag that can be used to show an instant, hashed information of the beam and machine, a proposal based on the beam lifetime study was arranged. After an off-line analysis based on a series data from SSRF, the Touschek factor was confirmed to be sensitive to the beam sizes and other machine related parameters. It is also believed that this factor is able to reflect the change of the beam status fast enough.

The beam current data would be enough to calculate the Touschek factor so that it is very economic, simple and intuitively clear. Besides, the algorithm is convergence and needs little intervention, so that it is feasible to provide the Touschek factor to the operators or physicists as an on-line flag of the beam/machine status. Long-term indication capability had also been confirmed (as shown in figure~\ref{fig:TouschekFactor-TransverseArea-Manydays} that the daily data of the Touschek factors during a typical period of operations and the corresponding beam sizes, and the Touschek factor did not miss the changes of the beam size).
 
\begin{center}\begin{minipage}{\hsize}
\includegraphics[width=.8\hsize]{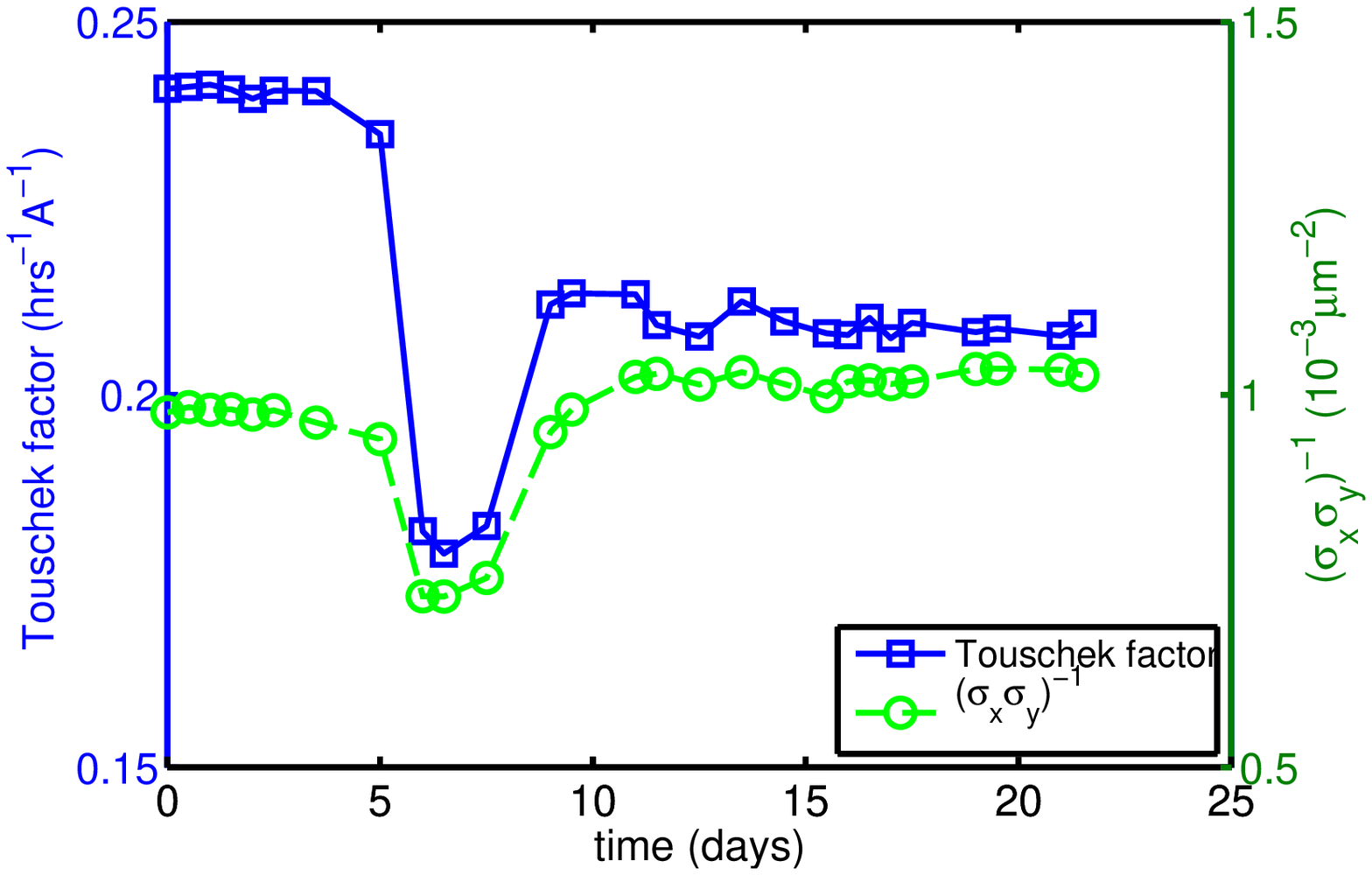}
\figcaption{\label{fig:TouschekFactor-TransverseArea-Manydays}Beam status tracking for long term operation.}
\end{minipage}\end{center}

During the experiments, the Touschek factor has shown some inspiring results in SSRF. Grouping the similar beam statuses could be useful for the operators when data were to be categorized before analysis. Finding the stepwise process of the transforming from one status to another could help the physicists during their beam experiments, such as looking for some critical parameters. Neither of them could be easily accomplished without the Touschek factor.

\bibliographystyle{ieeetr}
\bibliography{BI}

\begin{thebibliography}{10}

\bibitem{Wiedemann:PAP:2007}
H.~Wiedemann, {\em Particle Accelerator Physics}.
\newblock Springer-Verlag Berlin Heidelberg, 3rd~ed., 2007.

\bibitem{Khan:EPAC:1994}
S.~Khan, ``Simulation of the {T}ouschek effect for {BESSY} {II} a {M}onte
  {C}arlo approach,'' in {\em Proceedings of the Fourth European Particle
  Accelerator Conference}, (London, England), pp.~1192--1194, 1994.

\bibitem{Boscolo:PAC:2009:TH6PFP060}
M.~Boscolo, M.~Biagini, P.~Raimondi, M.~Sullivan, and E.~Paoloni, ``{T}ouschek
  background and lifetime studies for the {S}uper{B} factory,'' in {\em
  Proceedings of PAC09}, (Vancouver, BC, Canada), pp.~3844--3846, 2009.

\bibitem{Kang:APAC:2001}
H.~Kang, J.~Huang, and S.~Nam, ``Measurement of {T}oschek lifetime in {PLS}
  storage ring,'' in {\em Proceedings of the Second Asian Particle Accelerator
  Conference}, (Beijing, China), pp.~314--316, 2001.

\bibitem{Leonov:RuPAC:2004}
V.~Leonov and V.~Ushkov, ``Measurement of {S}iberia-1 beam lifetime by the
  decay rate plot method,'' in {\em Proceedings of RuPAC XIX}, (Dubna),
  pp.~272--274, 2002.

\bibitem{Steier:PAC:2009:TH5PFP033}
C.~Steier and L.~Yang, ``{T}ouschek lifetime measurements at small horizontal
  emittance in the {ALS},'' in {\em Proceedings of PAC09}, (Vancouver, BC,
  Canada), pp.~3269--3271, 2009.

\bibitem{Blinov:PAC:2011:MOP182}
V.~Blinov, V.~Kiselev, S.~Nikitin, I.~Nikolaev, and V.~Smaluk, ``Measurement of
  the energy dependence of {T}ouschek electron counting rate,'' in {\em
  Proceedings of 2011 Particle Accelerator Conference}, (New York, USA),
  pp.~426--428, 2011.

\bibitem{Nash:IPAC:2011:THPC008}
B.~Nash, F.~Ewald, L.~Farvacque, J.~Jacob, E.~Plouviez, J.~Revol, and
  K.~Scheidt, ``{T}ouschek lifetime and momentum acceptance measurements for
  {ESRF},'' in {\em Proceedings of IPAC2011}, (San Sebasti\'an, Spain),
  pp.~2921--2923, 2011.

\bibitem{Huang:NT:2010}
G.~Huang, J.~Chen, Z.~Chen, Y.~Leng, and K.~Ye, ``{X}-ray pinhole camera system
  design for {SSRF} storage ring,'' {\em Nuclear Techniques}, vol.~33,
  pp.~806--809, November 2010.

\bibitem{Chen:IPAC:2013:MOPME053}
Z.~Chen, Y.~Leng, J.~Chen, and G.~Huang, ``Point spread function study of
  {X}-ray pinhole camera in {SSRF},'' in {\em Proceedings of IPAC2013},
  (Shanghai, China), pp.~592--594, 2013.

\bibitem{Leng:IPAC:2011:TUPC116}
Y.~Leng, Y.~Yan, Z.~Chen, and R.~Yuan, ``Beam diagnostics global data warehouse
  implementation and application at {SSRF},'' in {\em Proceedings of IPAC2011},
  (San Sebasti\'an, Spain), pp.~1287--1289, 2011.

\end{thebibliography}
\end{multicols}

\end{document}